\documentclass[aip,jmp,amsmath,amssymb,floatfix,preprint]{revtex4-1}
\usepackage[english]{babel}
\usepackage{graphicx}  
\usepackage{dcolumn}   
\usepackage{bm}        
\usepackage{amssymb}   
\usepackage{amsmath}
\usepackage{verbatim}
\usepackage{enumerate}
\usepackage{mathtools}
\usepackage{xcolor}
\usepackage{hyperref}

\begin{document}

\title{Band alignment at semiconductor-water interfaces using explicit and implicit descriptions for liquid water}

\author{Nicolas G. \surname{H{\"o}rmann}}\affiliation{Theory and Simulation of Materials (THEOS), and National Centre for Computational
Design and Discovery of Novel Materials (MARVEL), \'Ecole Polytechnique F\'ed\'erale de Lausanne, 1015 Lausanne, Switzerland}
\author{Zhendong \surname{Guo}}\affiliation{Chaire de Simulation \`a l'Echelle Atomique (CSEA), \'Ecole Polytechnique F\'ed\'erale de Lausanne, 1015 Lausanne, Switzerland}
\author{Francesco \surname{Ambrosio}}\affiliation{Chaire de Simulation \`a l'Echelle Atomique (CSEA), \'Ecole Polytechnique F\'ed\'erale de Lausanne, 1015 Lausanne, Switzerland}
\author{Oliviero \surname{Andreussi}}\affiliation{Department of Physics, University of North Texas, Denton, TX 76207, USA}
\author{Alfredo \surname{Pasquarello}}\affiliation{Chaire de Simulation \`a l'Echelle Atomique (CSEA), \'Ecole Polytechnique F\'ed\'erale de
Lausanne, 1015 Lausanne, Switzerland}
\author{Nicola \surname{Marzari}}\affiliation{Theory and Simulation of Materials (THEOS), and National Centre for Computational Design
and Discovery of Novel Materials (MARVEL), \'Ecole Polytechnique F\'ed\'erale de Lausanne, 1015 Lausanne, Switzerland}

\date{\today}

\begin{abstract}
In this work we study and contrast implicit solvation models against explicit atomistic, quantum mechanical models in the description of the band alignment of semiconductors in aqueous environment, using simulations based on density functional theory. We find consistent results for both methods for 9 different terminations across 6 different materials whenever the first solvation shell is treated explicitly, quantum mechanically. Interestingly this first layer of explicit water is more relevant when water is adsorbed but not dissociated, hinting at the importance of saturating the surface with quantum mechanical bonds. Furthermore, we provide absolute alignments by determining the position of the averaged electrostatic reference potential in the bulk region of explicit and implicit water with respect to vacuum. It is found that the absolute level alignments in explicit and implicit simulations agree within $\sim 0.1-0.2$ V if the implicit potential is assumed to lie 0.33 V below the vacuum reference level. By studying the interface between implicit and explicit water we are able to trace back the origin of this offset to the absence of a water surface dipole in the implicit model, as well as a small additional inherent polarization across the implicit-explicit interface.
\end{abstract}


\maketitle

\section{Introduction}
Atomistic insight into the structure, composition and properties of solid-liquid interfaces is paramount for our understanding of the stability or performance of materials in many technological devices, such as chemical sensors, batteries or fuel cells. Very often, such detailed knowledge can only be obtained from atomistic simulations or from a combination of theory and experiments (e.g. interpretation of spectroscopic measurements).
To date, the main challenges for such simulations are the enormous size of the compositional and configurational space, in particular for \emph{electrochemical} solid-liquid interfaces, where the stability of a certain interface composition is not only influenced by the electrochemical potential of adsorbates but also by the energetics of the long-range space-charge layer, induced by the screening in the electrolyte over distances that are often out of reach for purely atomistic interface models due to the limits in cell size and number of atoms\cite{Hoermann2019}.

Recently, several studies have shown that coupling density functional theory (DFT) to implicit solvation models can yield accurate results for the solvation energies \cite{Andreussi2012,Letchworth-Weaver2012,Bonnet2013,Mathew2014,Ringe2017,Hoermann2019,Nattino2019}. In these cases, the explicit electrolyte solution (e.g. water with ions) is substituted by a mean-field description of the solvent, where the liquid is modelled via a polarizable continuum or via joint density-functional theory\cite{Jinnouchi2008, Andreussi2012, Fisicaro2016,Letchworth-Weaver2012,Mathew2014, Ringe2017}. Computational costs are highly reduced not only because of smaller quantum mechanical systems but also because thermodynamic averaging, which is mainly necessary for the correct description of the liquid, is substituted by the response of the continuum model. 

Reasonable agreement between explicit and implicit descriptions of water as well as with experiment also typically found for first principles simulations of metallic slabs with respect to interfacial structure, capacitance, potential of zero charge and interface energetics\cite{Dabo2008,Dabo2010,Letchworth-Weaver2012,Bonnet2013,Mathew2014,Sakong2015,Sakong2016, Hansen2016a, Sundararaman2017, Hoermann2019, Gauthier2019}. This points to rather unspecific interfacial water structures. Studies have shown that water ordering on metals depends mainly on the relative size of surface-water and water-water interactions\cite{Schnur2009,Gross2014,Biswas2018}, where significant ordering occurs only for very strong or very weak surface-water interactions, where the surface acts as a template or as a mere boundary supporting 2-dimensional intermolecular ordering via H-bonds.

\begin{figure}
\centering
\includegraphics[width=0.6\columnwidth]{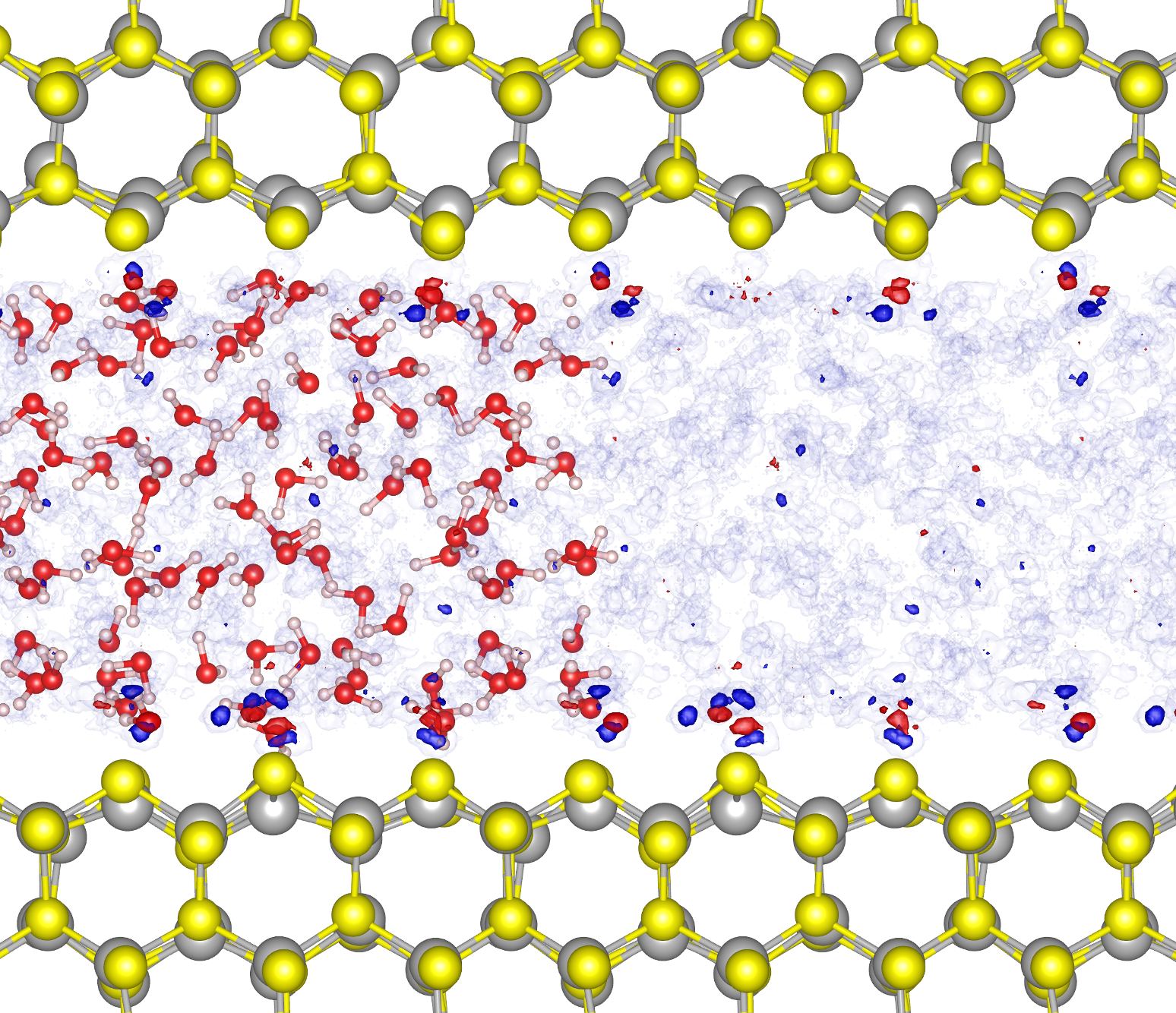}
  \caption{CdS(10$\bar{1}$0) slab in a liquid water environment. (left) AIMD snapshot. (right) Time average (8.6 ps) of O (red) and H (blue) densities, with low (semitransparent) and high (solid) density isosurfaces. As evidenced by density maxima close to the CdS surface (solid red/blue isosurfaces),  interfacial water molecules are less mobile than more distant water molecules.}
  \label{fig:water_MD_avg}
\end{figure}

For multicomponent semiconductor-water interfaces, on the other hand, rather strong surface-water interactions are more common\cite{Kerisit2005,Feibelman2010, Zemb2011, Parsegian2011, Bjoerneholm2016, McBriarty2017, McBriarty2018,Monroe2018} inducing very interface-specific properties of the solvation shell, where the applicability of implicit solvation is yet to be tested. As an example, we plot the time averaged density of O and H atoms as obtained from an ab-initio molecular dynamics (AIMD) trajectory of CdS(10\=10) in explicit water in Fig. \ref{fig:water_MD_avg}. The inhomogeneous accumulation of O and H (red and blue isosurfaces) at the interface are clear signatures of immobile, interfacial H$_2$O molecules, whose properties are expected to be significantly different from those of bulk water molecules.

In this work we test protocols to simulate the band alignment of semiconductors accurately leveraging implicit solvation models, and validate them against explicit simulations for the following semiconductors in water: (rutile) r-TiO$_2$(110) and CdS(10$\bar{1}$0) with molecularly adsorbed water, GaN(10$\bar{1}$0) with dissociatively adsorbed water and a-TiO$_2$(101), GaAs(110) and GaP(110) with either of these two possibilities. All these terminations were previously found to be stable or meta-stable\cite{Guo2018}. We will show that it is necessary to describe the first solvation shell explicitly in order to capture interfacial potential drops at semiconductor-water interfaces accurately within an implicit model, as similarly argued in Refs. \onlinecite{Ping2015,Blumenthal2017}. Furthermore, we will provide an estimate for the position of the flat electrostatic potential in the bulk of the implicit model that is approximately 0.33 eV below the vacuum level (absolute potential scale) for the SCCS implicit solvation model\cite{Andreussi2012}. This work provides best practices for simulations of semiconductor-water interfaces in implicit aqueous environment and also new insights in the effects and properties of interfacial water layers and to what extent they can be described by coarse grained continuum models, which we will tackle in the near future.

\section{Method}
\begin{figure}
\centering
\includegraphics[width=0.6\columnwidth]{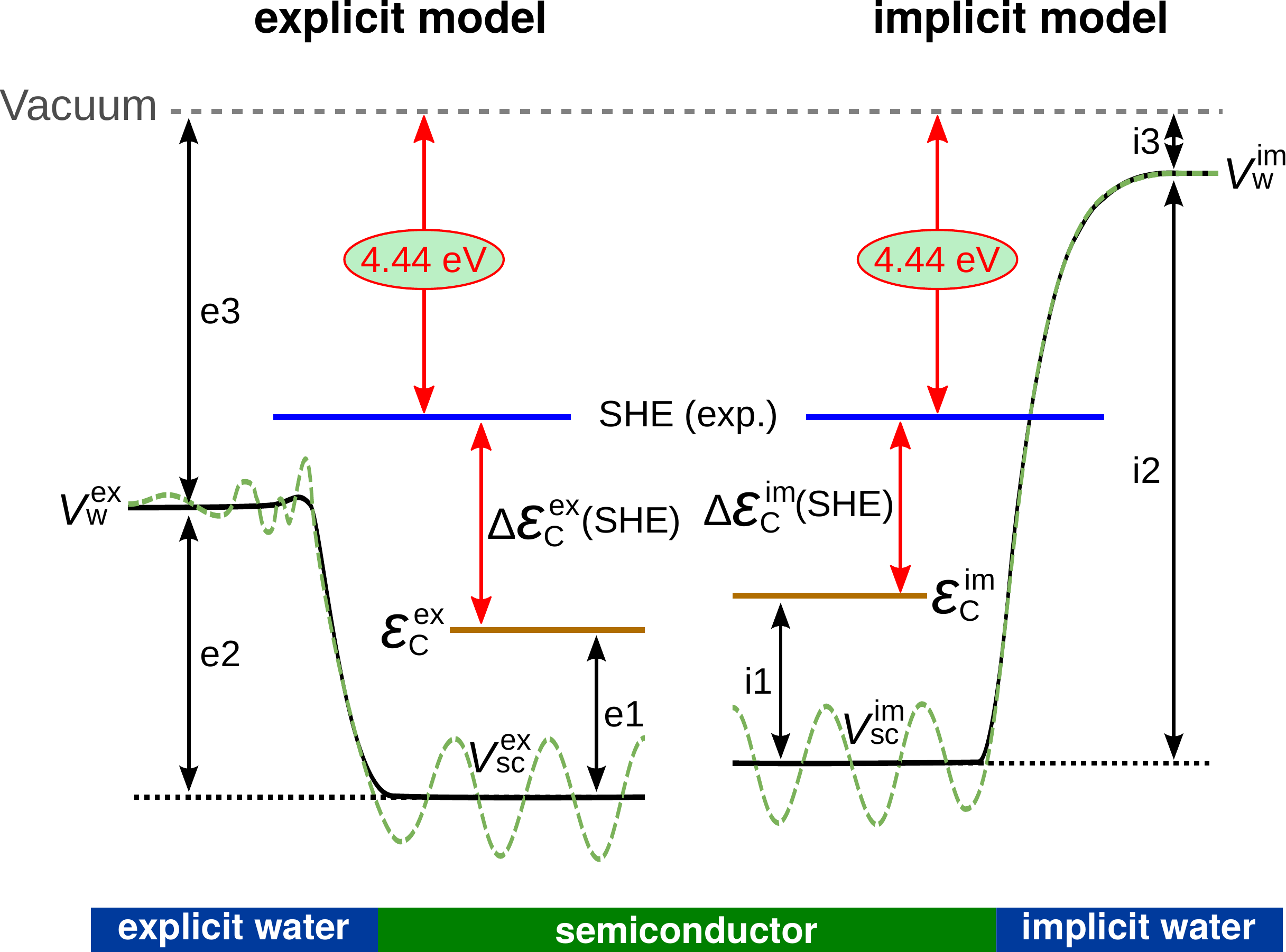}
\caption{Schematics of band-alignment at semiconductor-water interfaces using explicit and implicit descriptions for liquid water. For both models, three separated calculations marked as \textbf{e1}-\textbf{e3} and \textbf{i1}-\textbf{i3} are needed. In order to directly compare the band alignment results achieved through these two models, the vacuum level (gray dashed) and experimental standard hydrogen electrode (SHE) (blue solid) are selected as the common reference, with the latter being 4.44 eV lower than the former. The electrostatic potentials across the interfaces are shown in green dashed and solid black curves. The conduction band edges of the semiconductors ($\epsilon_{\rm C}^{\rm ex}$/$\epsilon_{\rm C}^{\rm im}$, brown lines) are shown with respect to the average electrostatic potentials ($V_{\rm sc}^{\rm ex}$/$V_{\rm sc}^{\rm im}$) of the bulk semiconductors. $V_{\rm w}^{\rm ex}$/$V_{\rm w}^{\rm im}$ represent the average electrostatic potentials in the regions of water far away from the interfaces.}
\label{fig:ES_reference_potentials}
\end{figure}

The approaches adopted for the determination of band alignment at semiconductor-water interfaces are schematically illustrated in Fig. \ref{fig:ES_reference_potentials}. For both interface models, explicit (\textbf{e}) and implicit (\textbf{i}), we simulate an explicit semiconductor slab with potential adsorbates (e.g. explicit water molecules) immersed in an explicit aqueous or an implicit environment. In order to determine the band alignment on an absolute scale (e.g. w.r.t. vacuum or SHE) three potential offsets need to be known: They are marked as e1-e3/i1-i3 in Fig. \ref{fig:ES_reference_potentials} and correspond to the position of the bands, e.g. the bottom of the conduction band $\epsilon_{\rm c}$, w.r.t. the average bulk electrostatic potential $V_{\rm SC}$ (e1/i1), the alignment of $V_{\rm SC}$ w.r.t. the average bulk water potential $V_{\rm W}$ (e2/i2) and the position of $V_{\rm W}$ w.r.t. an absolute reference, most conveniently vacuum (e3/i3). 

Whereas bulk simulations \textbf{(e1/i1)} are equivalent for explicit and implicit models, the alignment of $V_{\rm SC}$ with respect to the water average $V_{\rm W}$ \textbf{(e2/i2)} will be different. This is due to the fact that the main electrostatic contribution of the average potential in explicit water - the quadrupole contribution \cite{Leung2010, Ambrosio2018_JPCL} - is missing in typical implicit models and due to different interface models used, e.g. with varying amount of explicitly treated interfacial water. In particular, it is yet unclear which and how many water molecules to treat explicitly, and whether similar results as in explicit simulations can be obtained. In order to answer both of these questions we analyse the properties of individual water molecules and interfacial water layers from all-explicit ab-initio molecular dynamics (AIMD) simulations for the 9 aforementioned water-semiconductor interfaces. Based on these results, we construct representative sampling sets of interface models with 0 to 3 explicit interfacial water layers, by removal of appropriate water molecules and using 100-150 random snapshot of the all-explicit AIMD trajectories per studied interface. Subsequently, the potential offsets are determined by SCF calculations based on density functional theory from the average electrostatic potential alignment $V^{\rm ex}_{\rm SC}$ w.r.t. $V^{\rm ex}_{\rm W}$ for the all-explicit snapshots and $V^{\rm im}_{\rm SC}$ w.r.t. $V^{\rm im}_{\rm W}$ for the interfaces with reduced number of molecules embedded in an implicit environment.

It will be shown that the potential offset $V^{\rm ex}_{\rm W}-V^{\rm im}_{\rm W}$ is independent of the interface termination and amount of explicit water beyond the first solvation shell, supporting the suggestion that $V^{\rm im}_{\rm W}$ can indeed be used as a universal reference level.
For the final determination of the alignment with respect to an absolute level (vacuum) \textbf{(e3/i3)}, $V^{\rm ex}_{\rm W}$ is determined by simulation of the alignment of the potential of an explicit water slab in vacuum. For the implicit model we provide two pathways, namely by a fit to reproduce all-explicit calculations as well as a dedicated study of the potential alignment of explicit water slabs in implicit water, with both leading to consistent results.

It is important to note that all potential offsets (1-3) are not necessarily transferable between different DFT codes or pseudopotentials; however, the alignment of energy levels of the band edges w.r.t. vacuum remains unaltered, due to a cancellation in the differences (see also the comment below in section \ref{sec:comp_setup}). 

\section{Computational Setup}
\label{sec:comp_setup}
The fully equilibrated molecular dynamics trajectories used in this work were obtained in Ref.\ \citenum{Guo2018} and Ref.\ \citenum{Ambrosio2018_JPCL} by Car-Parrinello AIMD simulations in the NVT ensemble at 350 K with a time step of 0.5 fs and using a liquid water-adapted rVV10 functional\cite{Miceli2015}, that accounts for nonlocal van der Waals interactions\cite{Vydrov2010,Sabatini2013}. Timeframes were carefully chosen for the 9 selected interface terminations such that no drifts of potential or change of interfacial composition are present in the trajectories. For each interfacial system a thermal distribution of structures was approximated by a random selection of 100-150 snapshots from a subset of structures corresponding to each 50th MD step ($\Delta t$ = 25 fs). 
All reported results here are based on averages, obtained from SCF calculations of these snapshots and derived structures using the ENVIRON\cite{quantum-environment} module of Quantum ESPRESSO\cite{Giannozzi2009} and PBE\cite{Perdew1996} as exchange-correlation functional. As we are mainly interested here in demonstrating and testing the consistency of calculations for potential offsets at interfaces within implicit solvation models, we restrict the analysis to PBE, which underestimates band gaps but has been shown to reproduce potential offsets in good agreement with more advanced functionals and methods\cite{Guo2018}. We use pseudopotentials from the SSSP library\cite{Prandini2018} (v0.7, PBE, efficiency) with density and wavefunction cutoffs of 45 and 360 Ry, respectively and $\Gamma$-point-only sampling for the MD snapshots with a minute cold smearing\cite{Marzari1999} of 0.001 Ry. The simulations were managed with the materials' informatics infrastructure AiiDA \cite{Pizzi2016}. 

For the implicit model we use the SCCS formulation of the dielectric cavity\cite{Andreussi2012,Dupont2013,Andreussi2014} as implemented in ENVIRON with the chosen meta-parameters tuned for correct solvation energetics of explicit H$_2$O in the implicit model (\texttt{env\_static\_permittivity = 78.3, env\_pressure = -0.35 GPa, env\_surface\_tension = 50 dyn/cm, rhomax = 0.005 a.u., rhomin = 0.0001 a.u.}). As recently shown\cite{Andreussi2019}, it is necessary to construct a \emph{solvent-aware} SCCS boundary in order to prevent artificial dielectric pockets. Here, we use the optimized threshold value of 0.75 (see Ref. \onlinecite{Andreussi2019} and table ST3 in the SI). Calculations in slab geometry with implicit regions are typically non-symmetric and exhibit artificial effects due to periodic boundary conditions. These biases are counteracted by  using the dipole correction and extending the list of calculations used for final analysis with mirror symmetric copies of structures and derived quantitities e.g. electrostatic potentials.
It is important to note here that the charge density of atomic nuclei is smeared out in ENVIRON which renormalizes the total quadrupole moment of the simulation cell and thus the average electrostatic potential. As a result, electrostatic potentials and electronic level positions are different from the standard Quantum ESPRESSO results, and the reported potential offsets (1-3) are ENVIRON-specific and not necessarily transferable, which is also true for results obtained with other codes or pseudopotentials. Yet, all observable quantities (e.g. band alignment of slabs in vacuum) will be consistent across codes and pseudopotentials when preformed correctly (see also the discussion in section B of the SI).

\section{Results}
\subsection{Step e1/i1: Bulk levels}

PBE band gaps, and band alignment with respect to the average electrostatic potential were calculated for the bulk materials using the lattice constants of Ref. \onlinecite{Guo2018} and a k-point sampling with Monkhorst-Pack meshes corresponding to a k-point distances of $\leq$ 0.2 $\AA^{-1}$ (see section B in the SI).

\subsection{Interfacial water properties and interface model construction}
\label{sec:interfacial_water}

\begin{figure}
\centering
\includegraphics[width=0.6\columnwidth]{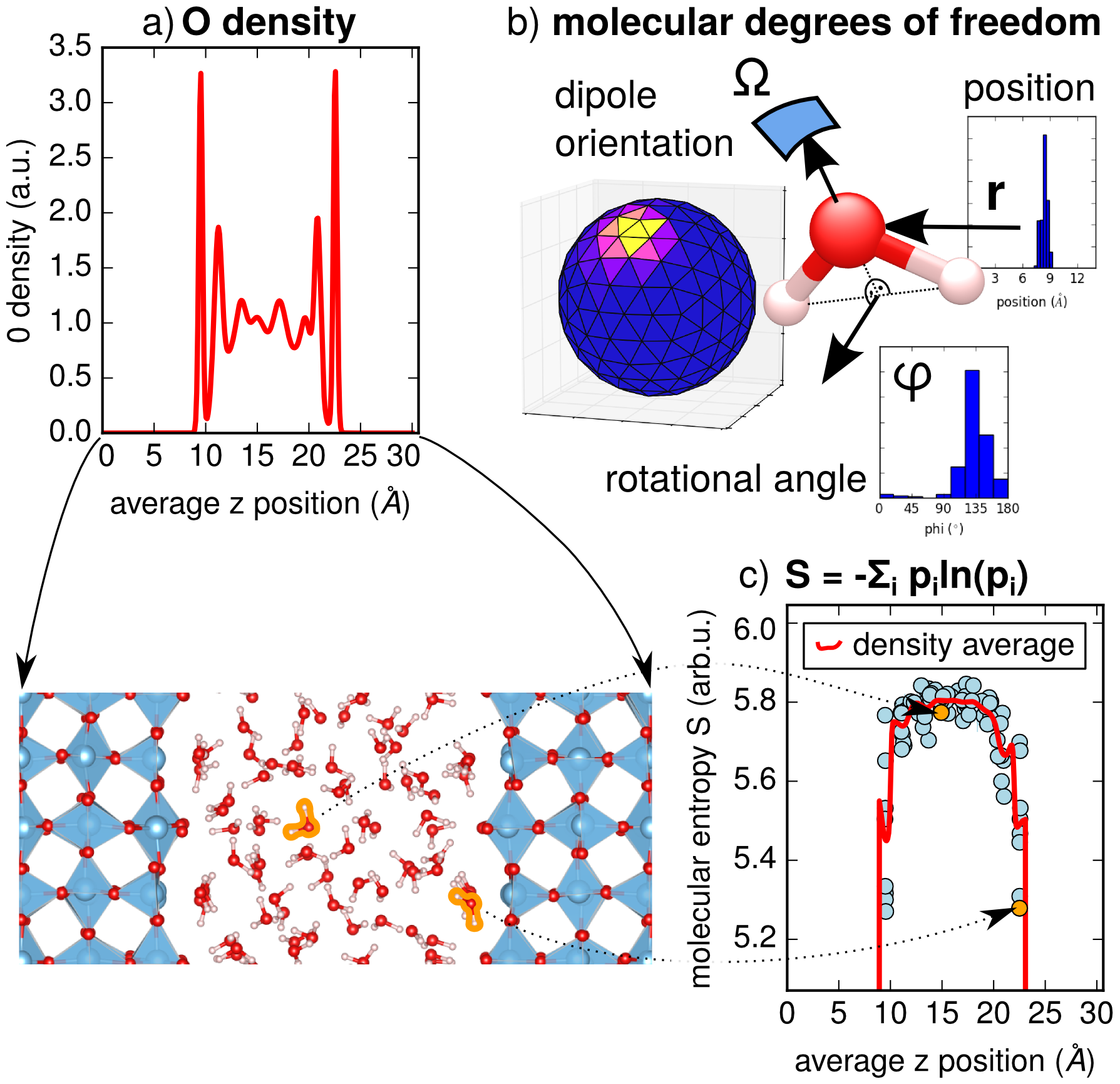}
  \caption{Interfacial water properties for r-TiO$_2$. a) Time-averaged oxygen density for water-related H$_2$O. b) Visualization of the studied molecular degrees of freedom (the position \textbf{r}, the orientation of the molecular dipole $\Omega$ and the rotational alignment $\phi$ of the normal to the H-O-H plane) and constructed histograms. Unbiased analysis of orientations was obtained by decomposing the spherical surface into 320 equal, equilateral triangles. c) Individual and density-averaged molecular entropies $S=-\sum_i p_i \ln p_i$ indicate distinct properties only for the first solvent shell. (See text and the SI for the definition of $p_i$.)}
  \label{fig:water_degrees_of_freedom}
\end{figure} 

\begin{figure}
\centering
\includegraphics[width=0.6\columnwidth]{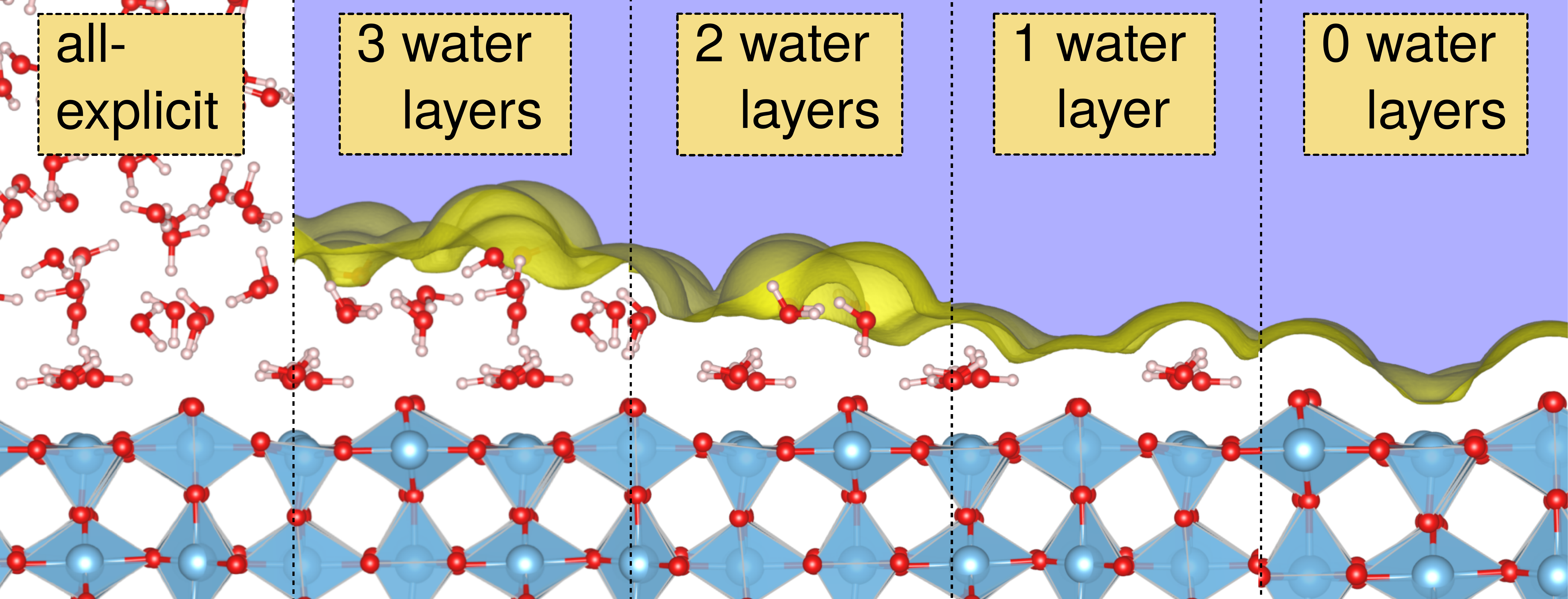}
  \caption{Illustration of the r-TiO$_2$-water interfaces studied with non-dissociatively adsorbed H$_2$O for the all-explicit and implicit descriptions of water with different number of interfacial water layers treated quantum mechanically. Regions filled with implicit dielectric are highlighted in light blue. The number of H$_2$O per water layers is defined by the first density peak of the water-related oxygen (see Fig. \ref{fig:water_degrees_of_freedom} a). }
  \label{fig:slabs_layers}
\end{figure}

Surface-water interactions are generally very strong for semiconductors, such that theoretical understanding of hydration is typically separated into properties of primary and secondary hydration shells\cite{Bjoerneholm2016,Zemb2011}, where the latter is dominated by generic water-water interactions more amenable to continuum models. In contrast, the primary hydration shell (2-4 \AA) with the first layer of adsorbed H$_2$O is very surface-specific, as it is dominated by enthalpic interactions with the solid surface which can vary by orders of magnitude \cite{Feibelman2010,Parsegian2011}. 

The distinct behaviour of molecules within the first solvation shell is evidenced in the systems studied here by the increased localization of water molecules in direct vicinity of surfaces, e.g. for CdS in Fig. \ref{fig:water_MD_avg}. 
In order to obtain a more quantitative picture we performed a combined analysis of cumulative and molecule-specific descriptors, in particular the planar average of the water-related oxygen density (Fig. \ref{fig:water_degrees_of_freedom} a) and a molecular entropy descriptor $S$ (Fig. \ref{fig:water_degrees_of_freedom} c), with:

\begin{equation}
    S = -\sum_i p_i \ln p_i\quad;\quad \sum_i p_i = 1
\end{equation}
The probabilities $p_i$ describe the sampling of the degrees of freedom of the H$_2$O molecule and are estimated from histograms with bins $i$ in the 6-dimensional molecular configuration space as visualized in Fig. \ref{fig:water_degrees_of_freedom} b): the position in space \textbf{r}, the orientation of the molecular dipole $\Omega$ and the rotational alignment $\phi$ of the normal to the H-O-H plane. 
We note in passing that choosing anisotropic bins for the orientation $\Omega$ - e.g. in spherical coordinates with ${\rm d}\Omega = \sin\theta {\rm d}\Phi{ \rm d}\theta$ - leads to artificial anisotropies, which is why we decided instead to use as bins 320 equal, equilateral triangles of an accordingly decomposed spherical surface (see Fig. \ref{fig:water_degrees_of_freedom} b). 
More details are discussed in section C of the SI. $S$ can be used to characterize the properties of water molecules, where small numerical values correspond to less mobile, more strongly bound H$_2$O as it relates to narrower probability distributions $p$. Due to limited simulation time scales (4-10 ps) major restructurings within the water slabs are absent, i.e. interfacial water molecules are not exchanging place with bulk-like water. Hence, molecules can be associated with individual water layers and used as probes for layer-specific properties. 

According to the molecular entropy-based analysis - as shown clearly for r-TiO$_2$ with molecularly adsorbed water in Fig. \ref{fig:water_degrees_of_freedom} c) - only the molecules of the first water layer - if associated with the first, pronounced oxygen density peak (Fig. \ref{fig:water_degrees_of_freedom} a) - show a distinctly different behavior than bulk-like water in the center of the water slab. H$_2$O molecules in the additional layers, which become increasingly ambiguous due to reduced density oscillations, are found to be not significantly different from bulk ones (Fig. \ref{fig:water_degrees_of_freedom} a). Similar results are found for the other systems (see Figs. SF4 - SF8 in the SI), with the first density peak being significantly more pronounced for the unpassivated surfaces (i.e. the pristine surfaces with molecularly adsorbed water). 

As a result, we chose to use the number of water molecules associated with the first density peak $N_l$ (for non-dissociated water) to define a water layer. In this setup, $n$ water layers correspond to the collection of the $n \times N_l$ water molecules closest to the surface. In most cases, but not always, this coincides with minima in the observed O-density and is associated with a "discontinuous jump" in the average distance to the surface between the $n \times N_l$th and the $n \times N_l + 1$st water molecule. Typical interfacial structures for r-TiO$_2$ as studied below are plotted in Fig. \ref{fig:slabs_layers}. We note in passing that - although well defined in the simulations here - a separation of the first water layer and the oxide surface can be ambiguous\cite{McBriarty2017, McBriarty2018}. 

\subsection{Step e2/i2: Potential offsets at semiconductor/water interfaces}
\label{sec:step2}

\begin{figure}
\centering
\includegraphics[width=0.55\columnwidth]{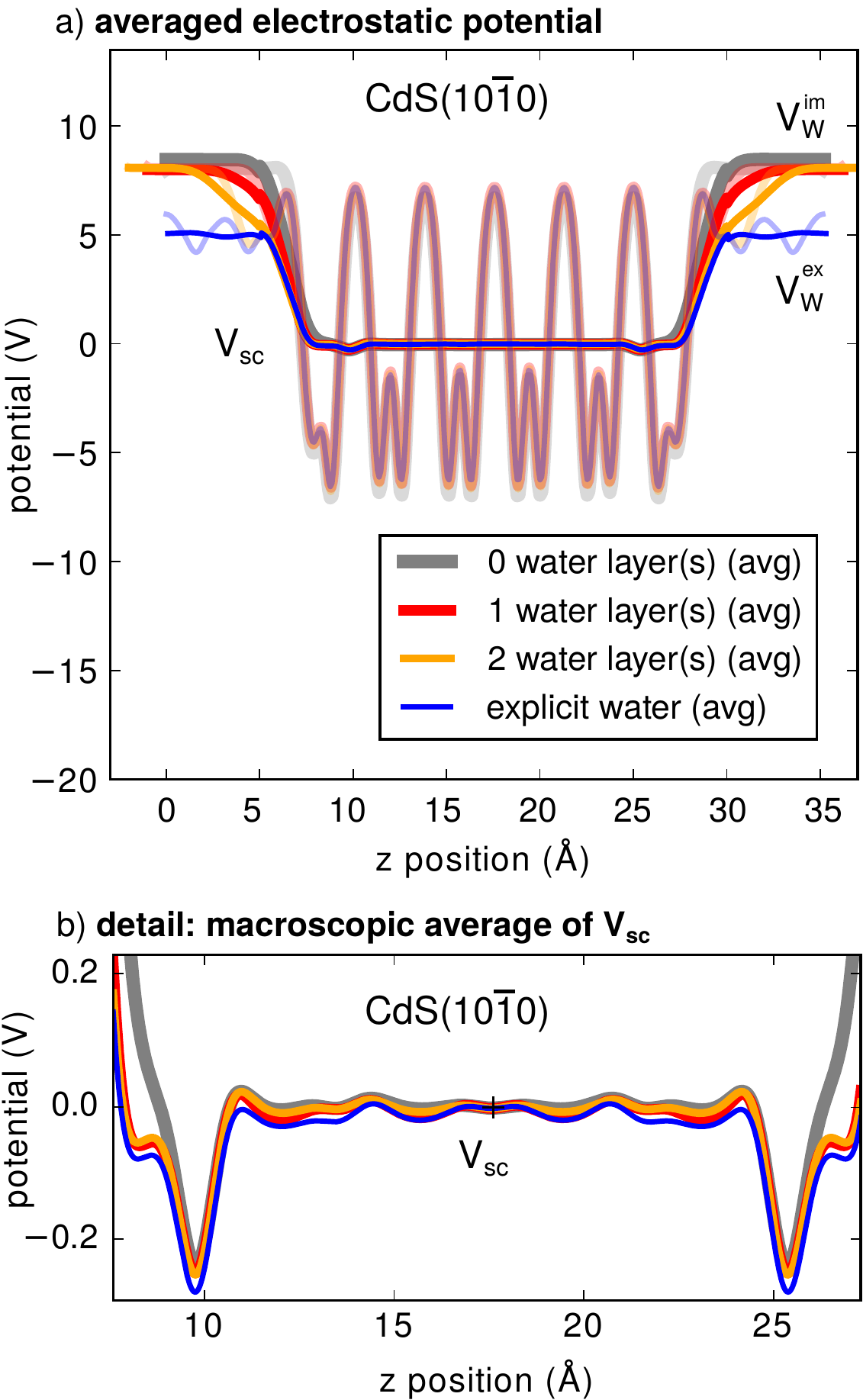}
  \caption{a) Planar ensemble average (semi-transparent) and macroscopic average (solid lines) of the electrostatic potential for CdS (10\=10) slabs in water. The macroscopic average of $V_{\rm SC}$ in the central 2 \AA of the slab is taken as a reference and set to 0 V. $V_{\rm W}$ is determined from the macroscopic average of the electrostatic potential in the center of the water region (implicit water: grey, red, orange with 0, 1, 2 explicit layers; explicit water: blue). b) Focus on the macroscopic average $V_{\rm SC}$ inside the semiconductor slab.}
\label{fig:V_w_CdS}.
\end{figure}

\begin{figure}
\centering
\includegraphics[width=0.6\columnwidth]{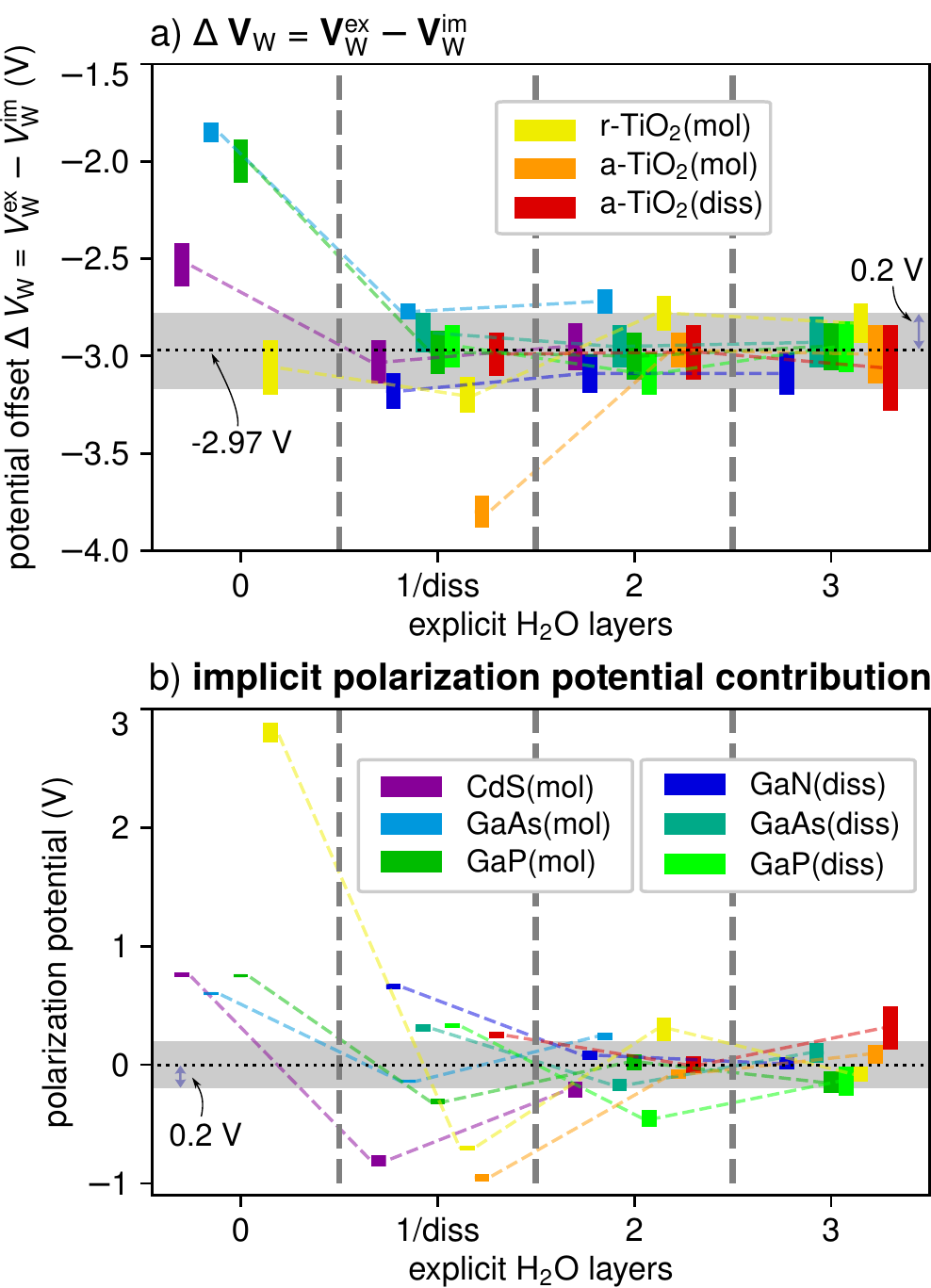}
  \caption{a) Offset of implicit and explicit water reference potentials $\Delta V_{\rm W} = V_{\rm W}^{\rm ex} - V_{\rm W}^{\rm im}$. $\Delta V_{\rm W}$ is nearly independent of the specific interface (material, number of explicitly treated water molecules) as long as a sufficient amount of water is treated explicitly (one or more explicit layers). The structure of the first water layer is indicated by the labels \emph{(mol)} (molecular water) and \emph{(diss)} (dissociated water); the potential offset of -2.97 V is the average for results beyond 1 explicit layer as discussed in section \ref{sec:step3}. b) Contribution to the potential offset from the polarization potential of the implicit model. Bar heights indicate the 95\% confidence interval. The grey horizontal bars are 0.4 V wide and are introduced as an aid to compare the different scales of panels a) and b).}
\label{fig:V_w_conv}.
\end{figure}

As discussed in the methods part, the average electrostatic potential inside bulk explicit water is qualitatively and quantitatively different from an implicit electrostatic potential due to the presence/absence of electronic and nuclear charge density. The hypothesis that both models are equally suited to determine band alignment can be tested by studying the potential offset of bulk explicit water and implicit water. For each interfacial system studied, the macroscopic average of the electrostatic potential in the center of the semiconductor slabs is used as reference point (e.g $V_{\rm SC}^{\rm im/ex}$ = 0), and the relative offset $\Delta V_{\rm W} = V_{\rm W}^{\rm ex} - V_{\rm W}^{\rm im}$ is determined. Averages and macroscopic averages of the electrostatic potential for a CdS (10\=10) slab in water are plotted in Fig. \ref{fig:V_w_CdS}.
Potential averages are calculated as planar and ensemble average of the electrostatic potential (semitransparent lines). The macroscopic averages are determined by applying an appropriately normalized, top-hat smoothing filter along the remaining z direction with a width corresponding to the central semiconductor layer. Inside the water region the filter is switched to being a smoothed top hat (combination of Gaussian error functions) with individually tuned width and smoothness, to give nearly constant water electrostatic potentials in the center of the explicit water region. In this way both the potential in the center of the semiconductor and in the fully explicit water region can be accurately averaged as they become essentially flat (see also section D in the SI). $V_{\rm W}$ corresponds to the value in the center of water and $V_{\rm SC}$ to the average in the central 2 \AA\ (marked as central cross in Fig. \ref{fig:V_w_CdS} c and SF9 - SF17 in the SI). Fig. \ref{fig:V_w_conv} a) summarizes the variation of the offset $\Delta V_{\rm W}= V_{\rm W}^{\rm ex} - V_{\rm W}^{\rm im}$ as a function of the number of explicitly treated water layers, where a dissociated water layer is counted as an explicit layer. The height of each bar corresponds to the uncertainty as given by the root of the summed squared errors of $V_{\rm W}^{\rm im}$ and $V_{\rm W}^{\rm ex}$ for a 95\% confidence interval (estimated by 1.96 $\times$ the standard error of the mean with ${\rm SEM} = \sigma/\sqrt{n}$; $\sigma$= the standard deviation, $n$= number of structures). $\Delta V_{\rm W}$ is found to be nearly independent of the specific material and number of explicitly treated water molecules as long as one or more layers of water are treated explicitly. In contrast, calculations with zero explicit water give significantly different and inconsistent results. This indicates that implicit simulations with at least one explicit water layer can yield consistent results with the all-explicit simulations, e.g. by application of a universal, relative shift of $\approx$-2.97 V, as suggested by the results of the next section (horizontal, dotted line in Fig. \ref{fig:V_w_conv} a). 

\begin{figure*}[t!]
\centering
\includegraphics[width=0.9\textwidth]{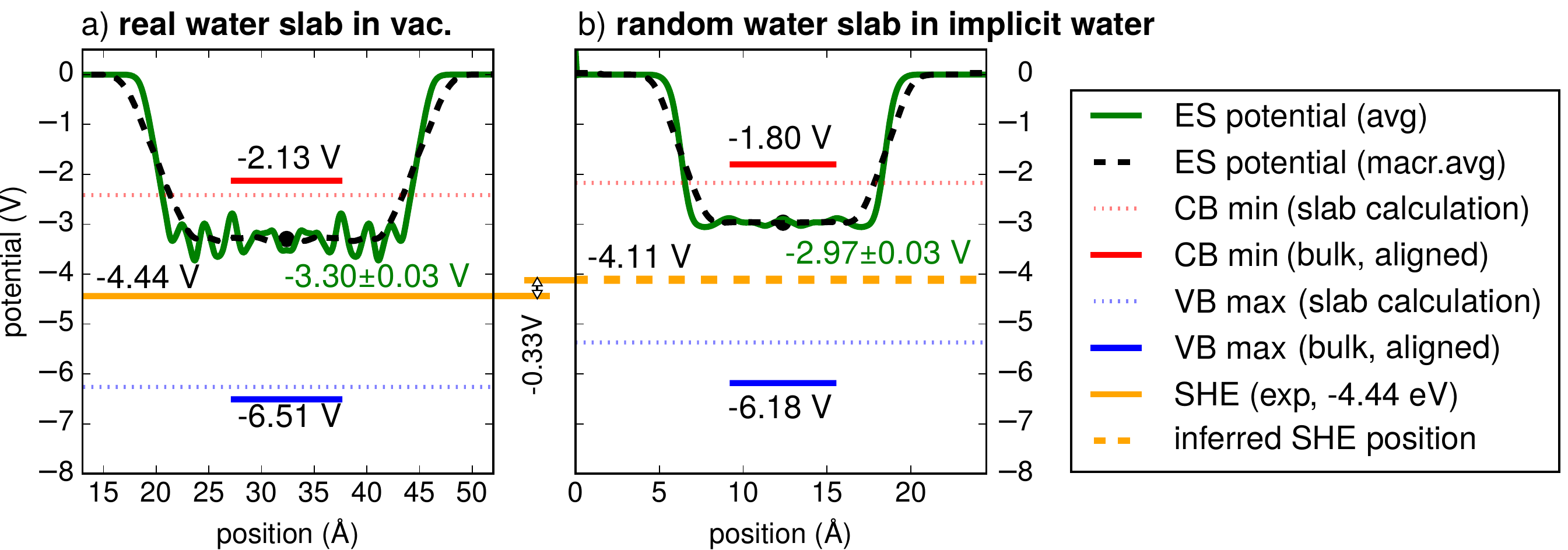}
  \caption{a) Absolute alignment of explicit water $V_{\rm W}^{\rm ex}$ from AIMD of a 25\AA\ thick water slab in vacuum (Ref. \onlinecite{Ambrosio2018_JPCL}) . b) The relative alignment of $V_{\rm W}^{\rm ex}$ for random water slabs (see text) in implicit solvation leads to a potential offset $\Delta V_{\rm W} = V_{\rm W}^{\rm ex} - V_{\rm W}^{\rm im} = -2.97$ V in perfect agreement with the results determined from averaging the offsets in Fig. \ref{fig:V_w_conv}. CB and VB edges are reported as averaged levels from the electronic states of the slab simulations and by introducing the respective bulk levels shifted to the ES average.}
  \label{fig:slabwater_alignment}
\end{figure*}

It is worth noting that anatase TiO$_2$ with molecularly adsorbed water is a problematic case for the implicit procedure. The SCF does not converge for a single calculation using the purely implicit model with the ENVIRON code also after tweaking typical parameters as reported in Ref. \onlinecite{Andreussi2019} and in private communications with other researchers (see also Figs. SF16 and SF17 in the SI). In a few cases it converges for a single explicit water layer, based on which we report averages. We speculate that this is due to surface Ti ions, whose electronic structure is extremely sensitive to the environment, making self-consistency more challenging. The reported results need to be taken with care due to possible biases for the small converged subset of calculations. Whether such convergence problems might universally hint towards the necessity of including explicit water beyond the first solvation shell is to be tested in the future.

\subsection{Step e3/i3: Relative and absolute alignment of explicit and implicit water}
\label{sec:step3}

It follows from the results in the previous section that $V_{\rm W}^{\rm im}$ can be used in the same way as $V_{\rm W}^{\rm ex}$ as a potential reference. In particular, if $V_{\rm W}^{\rm ex}$ is known on an absolute scale $V_{\rm W}^{\rm im}$ can be determined by a shift with $\Delta V_{\rm W}$. The absolute position of $V_{\rm W}^{\rm ex}$ is determined most conveniently from simulations of explicit water slabs in vacuum (trajectories already used in Ref. \onlinecite{Ambrosio2018_JPCL}) and found to be -3.30 V here (Fig. \ref{fig:slabwater_alignment} a and table \ref{tab:pot_offset_water}). The apparent difference to results from Ref. \onlinecite{Ambrosio2018_JPCL} (-3.68 V vs -3.30 V here) is due to different pseudopotentials used and the discussed ENVIRON-specific electrostatic (ES) potentials with smeared nuclei.

\begin{table}[]
    \centering  
    \caption{Position of the macroscopic average of the electrostatic potential $V_{\rm W}^{\rm ex}$ in the center of explicit water slabs w.r.t. the vacuum or the implicit solvation potential (SCCS) (c.f. Fig. \ref{fig:slabwater_alignment}). 
    The system "real slab" refers to an AIMD trajectory a water slab in vacuum (Ref. \onlinecite{Ambrosio2018_JPCL}). The other systems, denoted by "random set", refer to the random water slabs created according to the prescription in the text. "Polarization pot." denotes the potential drop due to the polarization charges of the implicit model as e.g. plotted in Fig. \ref{fig:random_water_explanation} a).}
\label{tab:pot_offset_water}
    \begin{tabular}{|l|c|c|}
    \hline
   water slab systems  & $V_{\rm W}^{\rm ex}$ (V) & Polarization pot. (V)\\
   \hline
   vacuum: real slab & $-3.30\pm0.03$ &  - \\
   vacuum: random set 1 &  $-3.05\pm0.06$ & -\\
   vacuum: random set 2 & $ -3.01\pm0.08$ & - \\
   \hline
   SCCS: random set 1 & $-2.95\pm0.04$ & $-0.33\pm 0.06$  \\
   SCCS: random set 2  & $-3.02\pm0.05$ & $-0.33\pm 0.07$  \\
   SCCS: set 1+2 & $-2.97\pm0.03$ & $-0.33\pm 0.04$  \\
       \hline
    \end{tabular}
\end{table}

\begin{figure}
\centering
\includegraphics[width=0.6\columnwidth]{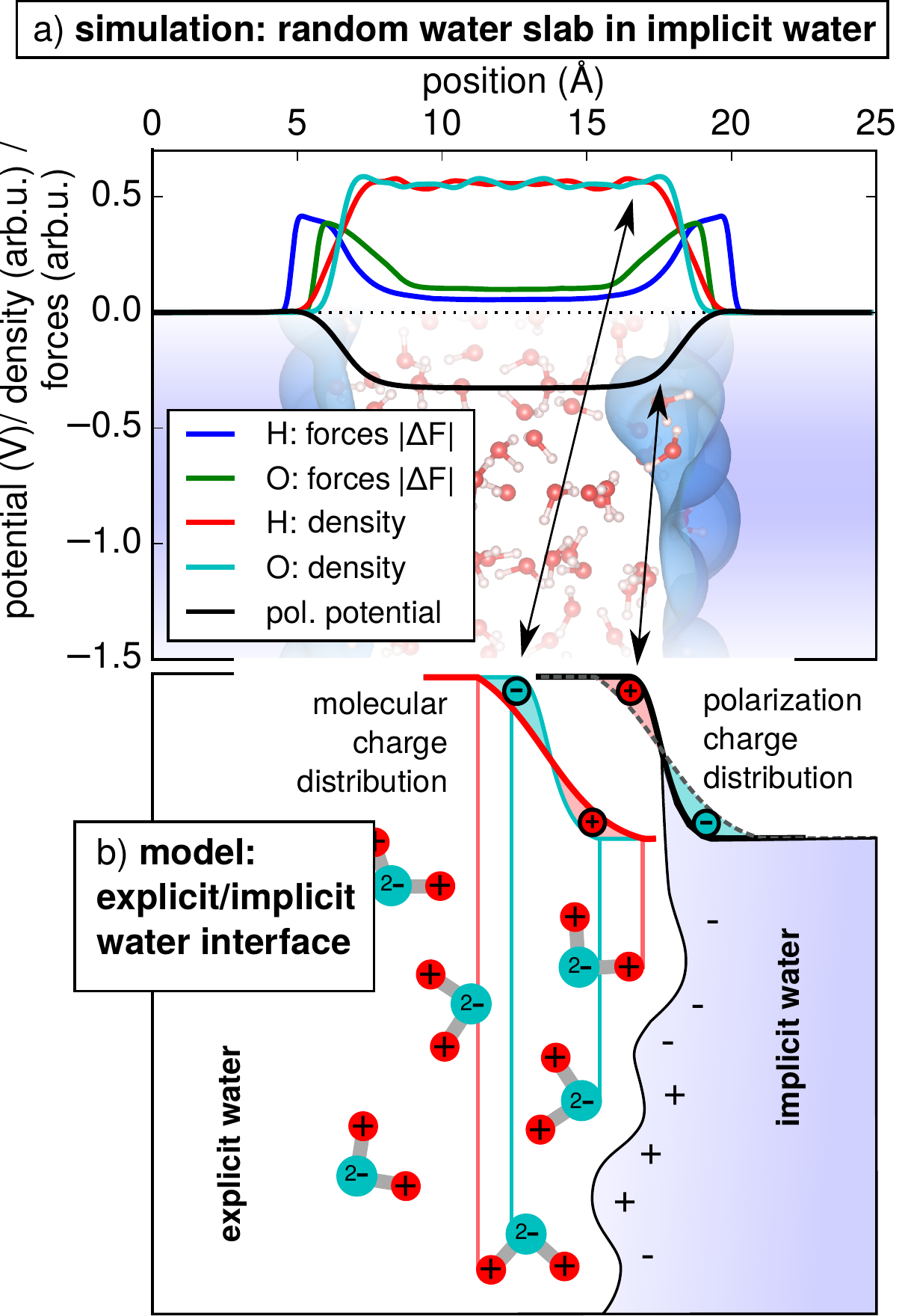}
  \caption{a) Properties of random explicit/implicit water interfaces (set 1+2): Atomic densities, force differences with respect to bulk calculations and planar average of the polarization potential. b) Explanatory model: The more smeared out H-density at the water surface as compared to the O-density (red and turquoise lines) leads to an inverted distribution of polarization charges at the explicit/implicit boundary and a polarization potential as observed in panel a.
  }
  \label{fig:random_water_explanation}
\end{figure} 

The average $\Delta V_{\rm W}$ from the simulations of Fig. \ref{fig:V_w_conv} a) with at least one explicit water layer is $2.97 \pm 0.05$ V; if the results with only one explicit layer are averaged (ignoring the results for a-TiO$_2$(mol)) the result is $3.00 \pm 0.10$ V. At the same time the polarization potential contribution to the overall electrostatic potential offset varies much more strongly than $\Delta V_{\rm W}$ across all interface types and number of explicit water layers (see Fig. \ref{fig:V_w_conv} a, b). This indicates that the implicit model can indeed reproduce the electrostatic response properties consistently across all studied explicit/implicit water interfaces and that the observed potential offset $\Delta V_{\rm W}\approx 2.97$ V might correspond in fact to a generic offset between implicit and explicit water.
To test this hypothesis, we constructed two random sets of explicit/implicit water heterostructures with random interfaces from a bulk water AIMD trajectory (as from Ref. \onlinecite{Ambrosio2018_JPCL}, 64 H$_2$O molecules, (12.42 \AA)$^3$). Set 1 consists of 205 'random' water slabs constructed by choosing time steps at random, the x,y,z lattice directions shuffled and flipped at random and a vacuum layer of 12.42 \AA\ introduced finally at a random position along the new z direction and respecting molecular integrity based on oxygen positions. Set 2 consists of 101 'random' slabs with a small total dipole moment for the water slabs, in order to see a potential influence of the long-range potential drops of polar slabs on the average potential alignment. More details on the properties of set 1 and set 2 are given in the SI, section E. The detailed results for the explicit/implicit potential offsets are reported in table \ref{tab:pot_offset_water}. We find negligible difference between set 1 and set 2 and thus combine both sets for our final reported offset of $-2.97 \pm 0.03$ V (Fig. \ref{fig:slabwater_alignment} b) in perfect agreement with the average of $\Delta V_{\rm W}$ from the results of the previous section. These results are confirmed by the extended analysis in the SI (section E) including 3 additional subsets of selected explicit/implicit heterostructures, supporting the generic origin of the observed offset.

Combining the alignment of explicit water in vacuum $V_{\rm W}^{\rm ex} = -3.30 \pm 0.03$ V, and the explicit/implicit offset $\Delta V_{\rm W} = -2.97\pm0.03$ V results in an absolute position of the implicit water potential of $V_{\rm W}^{\rm im} = -0.33 \pm 0.04$ V w.r.t. vacuum. As a result, if we position the standard hydrogen electrode (SHE) at -4.44 V, it is found at -4.11 V w.r.t. the implicit potential $V_{\rm W}^{\rm im}$. In order to understand better the origin of this offset in the implicit simulations, we decomposed the total electrostatic potential into quantum-mechanical contributions and polarization potential contributions. In addition, we also determined the alignment of the random water slabs in vacuum.
In agreement with the results of Ref. \onlinecite{Ambrosio2018_JPCL} the alignment of random water slabs in vacuum is slightly less negative as compared to the real water slab ($-0.15$ V as compared to $-0.25\pm0.07$ V here), which was interpreted as an effect of the water surface dipole\cite{Ambrosio2018_JPCL}. Along these lines of reasoning, we argue that the similar value of $V_{\rm W}^{\rm im} =-0.33$ V is mainly due to to the absence of a water surface dipole contribution in the implicit model. The additional difference of -0.08 V is due to an intrinsic implicit/explicit potential drop. However, it is worth noting that the polarization contribution from the implicit model is in fact significantly larger (-0.33 V, see table \ref{tab:pot_offset_water}), which indicates that changes of the interfacial electrostatic fields by the implicit model are largely cancelled by an opposite quantum mechanical charge redistribution. 

The origin of the implicit polarization for random water interfaces seems related to the structure of a 'random' water surface, as model charge distributions inspired by the observed H and O densities (red solid and turquoise lines in Fig. \ref{fig:random_water_explanation} a) can explain the observed potential drop (Fig. \ref{fig:random_water_explanation} b). They derive from the more smeared out H distribution within the water molecule and the interface construction protocol that imposes a hard-wall for the O density, and implicit polarization charges that follow similar spatial features but with opposite charges.

Fig. \ref{fig:random_water_explanation} a) also includes an analysis of force differences with respect to the periodic bulk calculations used for slab creation (red and green solid lines). The inclusion of mirror-symmetric results removes the dipolar potential drops across slabs in average, however, individually, most structures exhibit a macroscopic field across the slab inducing non-vanishing force differences $\vert\Delta F\vert$ in the center of the water slab (note O-related forces are approximately twice as large as H-related ones in the center of the slab, in agreement with electrostatic considerations). The largest force differences, however, are observed for water molecules directly at the explicit/implicit interface. How far such differences might induce differences between molecular dynamics results of pure explicit and explicit/implicit models is unknown, and will be studied in the future but it goes beyond the scope of this work.

\subsection{Semiconductor level alignment w.r.t. SHE}

\begin{figure*}[bth]
    \centering
    \includegraphics[width=1.\textwidth]{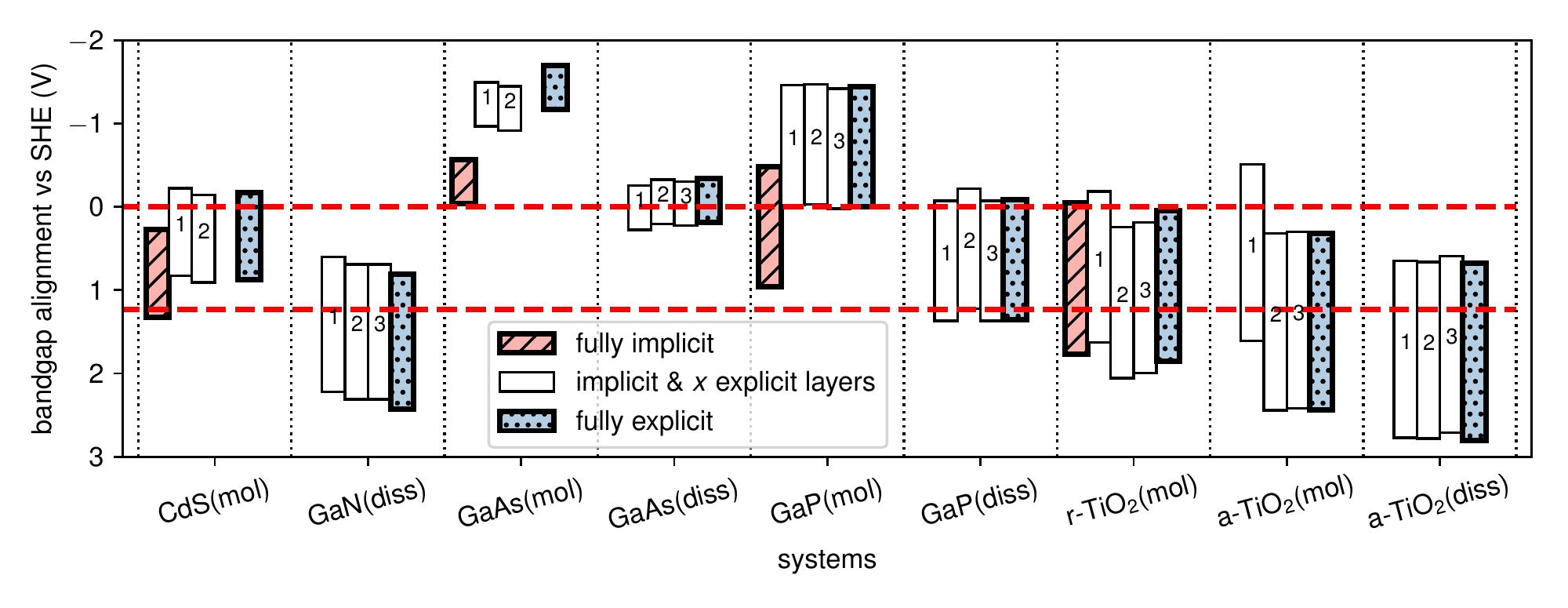}
    \caption{Band alignment of semiconductor levels in the different implicit/explicit water environments. The position of the band gap is plotted on the SHE scale, with the number of explicit water molecules increasing from left to right for each system. Systems without any explicit water are indicated by diagonal patterns, the all-explicit systems by dots. The technologically relevant potentials for hydrogen and oxygen evolution are plotted as dashed horizontal lines.}
    \label{fig:semiconductors_final_alignment}
\end{figure*}

With the above relative alignments and the SHE potential set to the experimental value at -4.44 V w.r.t. vacuum, it is possible to compare conduction and valence band alignments for all interface models and all simulated materials on an absolute scale, yielding Fig. \ref{fig:semiconductors_final_alignment}.

As mentioned before, apart from the case of a-TiO$_2$(mol) where results might be unreliable, no significant increase in accuracy for the potential alignment is observed when including water layers beyond the first solvation shell. The differences to the explicit systems are of the order of 0.1-0.2 V which is comparable to the accuracy of the sampling statistics. On the other hand, simulations without inclusion of explicit water might very likely exhibit alignment errors of 1 V or more which should be kept in mind in future band alignment studies, as this has significant impact on whether a system can support e.g. water splitting reactions or not. Note that we count the explicitly treated dissociated water layer as water layer. We think the observed pattern reflects a general principle that interfacial electrostatic potential drops are captured correctly whenever the QM-implicit interface exhibits no unsaturated dangling bonds which can influence closeby solvent molecules in a very specific way, not captured in implicit models.

\section{Summary and Discussion}
 In this study we showed that the band alignment of semiconductor slabs in the SCCS implicit model reproduces explicit results up to an accuracy of $\sim0.1-0.2$ V, provided that the amount of explicitly simulated water molecules is chosen appropriately. In particular, all systems considered, agree very well with fully explicit simulations, when the chemically interacting interface molecules or their fragments are treated explicitly, which corresponds to the first dissociated or undissociated water layer above the unpassivated surface, in agreement with the findings in Refs. \onlinecite{Ping2015,Blumenthal2017}. 
 This finding also agrees with the discussion in section C of the SI, where we observe that the unpassivated surface has interfacial water molecules with properties clearly different from bulk-like.

Furthermore, we studied the explicit-implicit water interface and showed that the electrostatic potential reference inside the implicit region does not correspond to the absolute potential reference (vacuum above the solution), but is shifted by approximately -0.33 V, which puts the SHE level at -4.11 eV w.r.t. the implicit level of the SCCS model. This offset is mainly related to the absence of an explicit water surface dipole contribution in the implicit model, and to a smaller extent to some generic polarization of the implicit model across the implicit/explicit interface.
The relative insensitivity of the overall level alignment to the materials and amount of explicit water with at the same time strongly varying implicit-model-related polarization contributions (Fig. \ref{fig:V_w_conv} and Figs. SF9 - SF17 in the SI) suggests that the SCCS model is indeed able to mimic accurately all generic solvent-related screening properties at explicit/implicit water interfaces. 

These results overall suggest that simulations of electrochemical interfaces can be performed with accuracies close to the all-explicit calculations at significantly reduced cost. We speculate that structural models with the QM-implicit boundary dominated by H, OH or H$_2$O species, as obtained with included explicit water might also be favourable for consistent energetics across different systems due to chemical and structural similarity.

The final step to develop an accurate simulation protocol for solid-liquid interfaces with implicit models will be to assess and validate ab-initio molecular dynamics simulations for mixed explicit/implicit models, something that we will address in a follow up paper.

\section{Acknowledgements}
The authors acknowledge partial financial support from the Swiss National Science Foundation (SNSF) through the NCCR MARVEL and the EU through the MAX CoE for e-infrastructure. This work has been realized in relation to the National. This work was supported by a grant from  the  Swiss  National  Supercomputing  Centre  (CSCS)  under  project  ID s836 and the computing facilities of SCITAS, EPFL.

\end{document}